\documentclass[aps,superscriptaddress,groupedaddress,twocolumn,floatfix,10pt]{revtex4}
\usepackage{natbib}
\usepackage{graphicx} % needed for figures
\usepackage{dcolumn} % needed for some tables
\usepackage{bm} % for math
\usepackage{amssymb,amsmath} % for math
\usepackage{textcomp} % for text
\usepackage[utf8x]{inputenc}
\usepackage{braket}
\usepackage{amsmath}
\usepackage{placeins}
\usepackage{xcolor}
\usepackage{pdfpages}
\usepackage{soul}
\usepackage{todonotes} %\todo[prepend/inline]{ }
\usepackage{changebar}
\usepackage{changes}
\definechangesauthor{AB}

\begin{document}

\pagestyle{empty}

\title{A bright future for financial agent-based models}

\author{J. Lussange(1), A. Belianin(2), S. Bourgeois-Gironde(3,4), B. Gutkin(1,5) \\, \\
1. Group for Neural Theory, LNC INSERM U960, Departement des Etudes Cognitives, Ecole Normale Superieure PSL University, Paris France \\ 
2.ICEF, National Research University Higher School of Economics and Primakov Institute for World Economy and International Relations, Russian Academy of Sciences \\
3.Institut Jean-Nicod, Ecole Normale Supérieure, Paris, FR \\
4.Institut d'Etude de la Cognition, PSL Research University, 75005, Paris, France\\
5.Center for Cognition and Decision Making, NRU Higher School of Economics, Moscow Russia \\
}

\date{%
%   1. Group for Neural Theory, LNC INSERM U960, DEC Ecole Normale Superieure PSL University, Paris France\\%
%   2. Organization 2\\
%   3. \\ 
%    4. Center for Cognition and Decision Making, NRU Higher School of Economics, Moscow Russia\\
    \today
}
\begin{abstract} 

The history of research in finance and economics has been widely impacted by the field of Agent-based Computational Economics (ACE). While at the same time being popular among natural science researchers for its proximity to the successful methods of physics and chemistry for example, the field of ACE has also received critics by a part of the social science community for its lack of empiricism. Yet recent trends have shifted the weights of these general arguments and potentially given ACE a whole new range of realism. At the base of these trends are found two present-day major scientific breakthroughs: the steady shift of psychology towards a hard science due to the advances of neuropsychology, and the progress of artificial intelligence and more specifically machine learning due to increasing computational power and big data. These two have also found common fields of study in the form of computational neuroscience, and human-computer interaction, among others. We outline here the main lines of a computational research study of collective economic behavior via Agent-Based Models (ABM) or Multi-Agent System (MAS), where each agent would be endowed with specific cognitive and behavioral biases known to the field of neuroeconomics, and at the same time autonomously implement rational quantitative financial strategies updated by machine learning. We postulate that such ABMs would offer a whole new range of realism. 

\end{abstract}

\maketitle

The field of finance and economics has used various approaches to model financial markets dynamics. Among these we can historically distinguish three important classes of models. The first and most encountered are statistical models which are calibrated to fit times series like past prices history. These can bring interesting results pertaining to general volatility~\cite{Bollerslev2008,Engle1982} or log-return forecasting~\cite{Brownlees2011} as long as the variability of the parameters of calibration is not too strong. The second are known as Dynamic Stochastic General Equilibrium (DSGE) models which provide explicit agent-based microfoundations for the sectoral dynamics and aggregate fluctuations~\cite{Sbordone2010}. Modern developments of DSGE models strive to add realism to the basic model structure, accounting for agent heteorgeneity, bounded rationality and imperfect learning, and (in the New Keynesian versions) replace the rational expectations hypothesis by market rigidities and exogenous stochastic shocks to emulate true market environment conditions~\cite{EvansHonka2001, Eusepi2011, Massaro2013}. These two classes of models have shown a variety of promising results over the years. However if we consider a top-down approach to system inference, we can say that they are based on rough approximations of reality~\cite{Farmer2009,Degrauwe2010}, and will not explain the wealth and diversity of price microstructure traditionally seen in markets. This leads to a third class of models called Agent-Based Models (ABMs) or sometimes Multi-Agent Systems (MAS) to probe and emulate markets from a pure bottom-up approach~\cite{Tesfatsion2006,Samanidou2007,Lebaron2002}, and considering them as the complex systems~\cite{Bonabeau2002} that they truly are. Among financial ABM models, we can also include order book models~\cite{Smith2003,Huang2015} even though some may see those as a midway approach. In a financial ABM, market investors or traders are modeled as agents trading together via an order book (such as a double auction order books~\cite{Mota2016}). This is a discrete-time algorithm taking in the trading bids at $t$ and offers of specific securities from all agents, and matching them at transaction prices which then collectively define the price of the market for such securities at the time step $t+1$. 

\vspace{1mm}

ABM have been used in many scientific disciplines~\cite{Macal2010,Axelrod1997,Grimm2006}. In economics, these models have emerged by way of psychological learning models~\cite{MostellerBush1955}, evolutionary biology~\cite{MaynardSmith1973, TaylorJonker1978}, and especially game theory~\cite{Mookherjee1994,ErevRoth1998,ErevRoth2014,CamererHo1999,FudenbergLevine1998}. In recent years, ABM also became popular as a tool to study macroeconomics~\cite{Colander2008,Dosi2013,Gualdi2016,Gualdi2015} --- specifically, the impact of trading taxes, market regulatory policies, quantitative easing, and the general role of central banks~\cite{Westerhoff2008}. ABMs can also play an important role in analysis of the impact of the cross-market structure~\cite{Xu2014}.

%\todo[inline]{\st{Learning in games at large uses ABM; Boris $to$ Alexis: can you please give an example and a couple of representative references. Can you please insert these directly in the document. I simply dont know them.}} 

From a regulatory point of view, this implies a general stronger role for ABM to play~\cite{Boero2015}. Jean-Claude Trichet declared for instance in $2010$: "\textit{As a policy-maker during the crisis, I found the available models of limited help. In fact, I would go further: in the face of the crisis, we felt abandoned by conventional tools. [...] Agent-based modeling dispenses with the optimization assumption and allows for more complex interactions between agents. Such approaches are worthy of our attention.}" 

\vspace{1mm}

Even if ABMs are often designed with many parameters and hence subject to the delicate issue of overfitting, one of their biggest advantages is that they require fewer assumptions (e.g normal distribution of returns, no-arbitrage) than top-down models~\cite{Lebaron2005}. Added to this, ABM display the famous complexity emergence proper to bottom-up approaches~\cite{Heylighen2008} and can hence show completely new regimes of transitions, akin to phase transitions in statistical physics~\cite{Plerou2003}. However, being models, ABMs are of course imperfect and need a thorough and lengthy cross-market validation~\cite{Hamill2016}. Yet at the same time, one should keep in mind that such a general and cautious validation of ABMs shall be in fact applicable and necessary to any other model as well~\cite{Wilcox2014}. 

\vspace{1mm}

From now on we consider the application of ABMs to financial markets. We shall note that among financial ABMs some exclusively pertain to high-frequency trading~\cite{Hanson2011,Bartolozzi2010}, while other take both high- and low-frequency into account~\cite{Wah2013,Paddrik2011,Aloud2013}. 
Another popular topic of literature in financial ABMs is the emulation of the widespread Minority Game~\cite{Challet2005,Demartino2006}, which formally is not a financial market problem, but a general game theory problem which can be related to the financial issues of pricing and forecasting. 

\vspace{1mm}

In order to generate a dynamic trading activity in financial ABMs, a basic economic assumption is that the agents disagree on the present security price or trade at different frequencies (which possibility is sometimes explicitly denied in economics literature~\cite{KyleObizhaeva2016}), and are hence willing to long or short a same security at different prices. In other words, there must be some sort of price disagreement happening and an original pricing mechanism at the discretion of each individual agent. In the literature, this mechanism of pricing in financial ABMs has in general been designed according to two basic mechanisms of trading: in some models at least a part of the agents trade in a random way as `noise traders'~\cite{Preis2006,Farmer2005,Maslov2000,Challet2003,Schmitt2012,Hanson2011,Bartolozzi2010}, and in other models agents use realistic trading strategies known to real financial markets, depending on the variability and stochasticity of the market~\cite{Lux2000,Cont2007,Bertella2014,Alfi2009}. 

\vspace{3mm}

\section{Accuracy}
%\todo[inline]{\st{Don't we need to have at least a sketch of an ABM model? We lack words, I suggest we do that, and place it here - before discussion of its advantages. Boris $to$ Alexis: I am not sure if we should go into technical details here. Describing an ABM in general vague terms is not useful, also to go into technical detail is also over kill and would take away from the generality of this opinion piece. We can always add a description if asked by the reviewers.} --- AB: Agree}

Over the years, economic research (and especially econophysics research) has gradually discovered a certain number of non-trivial statistical features of or stylized facts about financial times series. These stylized facts are based on variations in prices that have universal statistical properties in common from market to market, over different types of instruments, and time periods. Among these, those pertaining to returns distribution or volatility clustering for example were gradually discovered during the nineties: Kim-Markowitz~\cite{Kim1989}, Levy-Levy-Solomon~\cite{Levy1996,Levy1994,Levy1995,Levy1996b,Levy1996c,Levy1997,Levy2000}, Cont-Bouchaud~\cite{Cont2000}, Solomon-Weisbuch~\cite{Solomon2000}, Lux-Marchesi~\cite{Lux1999,Lux2000}, Donangelo-Sneppen~\cite{Donangelo2000,Donangelo2000b,Bak1999,Bak2001}, Solomon-Levy-Huang~\cite{Huang2000}. It was also not before this time that ABM started to emulate these stylized facts. 

\vspace{1mm}

The importance of the universality of stylized facts to really gauge financial markets comes from the fact that the price evolutions of different markets may have very different exogenous or endogenous causes. As a consequence they highlight general underlying financial mechanisms that are market-independent, and which can in turn be exploited for ABM architecture design. From a scientific point of view, stylized facts are hence extremely interesting and their faithful emulation has been an active topic of research in the past fifteen years or so~\cite{Lipski2013,Barde2015}. Their definite characteristics has varied ever so slightly over the years and across literature, but the most widespread and unanimously accepted stylized facts can in fact be grouped in three broad, mutually overlapping categories:

\vspace{1mm}

\noindent \textit{Non-gaussian returns}: the returns distribution is non-gaussian and hence asset prices should not be modeled as brownian random walks~\cite{Potters2001,Plerou1999}, despite what is taught in most text books, and applied in sell-side finance. In particular the real distributions of returns are dissimilar to normal distributions in that they are: (i) having fatter tails and hence more extreme events, with the tails of the cumulative distribution being well approximated~\cite{Cristelli2014,Potters2001} by a power law of exponent belonging to the interval $[2,4]$ (albeit this is still the subject of a discussion~\cite{Weron2001,Eisler2006} famously started by Mandelbrot~\cite{Mandelbrot1963} and his Levy stable model for financial returns), (ii) negatively skewed and asymmetric in many observed markets~\cite{Cont2001} with more large negative returns than large positive returns, (iii) platykurtic and as a consequence having less mean-centered events~\cite{Bouchaud1997}, (iv) with multifractal $k$-moments so that their exponent is not linear with $k$, as seen in~\cite{Ding1993,Lobato1998,Vandewalle1997,Mandelbrot1997}. 

\vspace{1mm}

\noindent \textit{Clustered volatilities}: market volatility tends to aggregate or form clusters~\cite{Engle1982}. Therefore compared to average, the probability to have a large volatility in the near-future is greater if it was large also in the near-past~\cite{Lipski2013,Devries1994,Pagan1996}. Regardless of whether the next return is positive or negative, one can thus say that large (resp. small) return jumps are likely followed by the same~\cite{Mandelbrot1963}, and thus display some sort of long memory behavior~\cite{Cont2005}. Because volatilities and trading volumes are often correlated, we also observe a related volume clustering. 
%\todo[inline]{Explicit reference to ARIMA}

%\vspace{3mm}

%\noindent \textit{Log-normal volatilities}: similarly to the distribution of price changes~\cite{Liu1999}, the cumulative probability distribution of the volatility defined as the local average of the absolute value of price changes over a given time window is well approximated by a power-law with exponent of around $3$. 

\vspace{1mm}

\noindent \textit{Decaying auto-correlations}: the auto-correlation function of the returns of financial time series are basically zero for any value of the auto-correlation lag, except for very short lags (e.g. half-hour lags for intraday data) because of a mean-reverting microstructure mechanism for which there is a negative auto-correlation~\cite{Cont2001,Cont2005}. This is sometimes feeding the general argument of the well-known Efficient Market Hypothesis~\cite{Fama1970,Bera2015} that markets have no memory and hence that one cannot predict future prices based on past prices or information~\cite{Devries1994,Pagan1996}. According to this view, there is hence no opportunity for arbitrage within a financial market~\cite{Cristelli2014}. It has been observed however that certain non-linear functions of returns such as squared returns or absolute returns display certain steady auto-correlations over longer lags~\cite{Cont2005}. 
%\todo[inline]{These three fact can be captured not only with ABM --- what are the advantages of it over other methods?}

\vspace{1mm}

Since then, ABMs of financial markets have steadily increased in realism and can generate progressively more robust scaling experiments. We can specifically highlight the potential of these simulations to forecast real financial time series via reverse-engineering. A promising recent perspective for such use of ABMs has been highlighted in the field of statistics by ~\cite{Wiesinger2012,Andersen2005,Zhang2013}: the agent-based model parameters are constrained to be calibrated and fit real financial time series and then allowed to evolve over a given time period as a basic forecast measure on the original time series used for calibration. With this, one could thus say that ABMs are now reaching Friedman's~\cite{Friedman1953} methodological requirement that a theory must be "\textit{judged by its predictive power for the class of phenomena which it is intended to explain}."

\vspace{3mm}

\section{Calibration}

Just as any other model, the parameters of the ABM must be calibrated to real financial data in order to perform realistic emulation. This part of calibration is together with architecture design the most technical and crucial aspect of the ABM~\cite{Canova2009}. Yet at the same time in the literature most calibration techniques are done by hand, so that the stylized facts are re-enacted in a satisfactory way. Therefore so far the ABM calibration step is often performed in way that is sub-optimal~\cite{Preis2006,Chiarella2009,Leal2016}. 

\vspace{1mm}

On the other hand, an efficient methodology for calibration would need two important steps. First a fully automated meta-algorithm in charge of the calibration should be incorporated, so that massive amount of financial data could be treated and the aforementioned scope of validity of ABM studied via cross-market validations~\cite{Fabretti2013,Preis2006}. This is important as the robustness of a calibration always relies on many runs of ABM simulations~\cite{Axtell2000}. Second, this calibration meta-algorithm should be working through the issues of overfitting and underfitting, which may constitute a severe challenge due to the ever-changing stochastic nature of financial markets. 

\vspace{1mm}

Part of this calibration problem is to thoroughly and cautiously define the parameter space. This step is particularly sensitive, since it can lead to potentially problematic simplifications. For instance, what should be the size of the time step of the simulation? ABM with a daily time tick will of course produce time series that are much coarser than those coming from real financial data, which include all sorts of intraday events~\cite{Gilli2003,Farmer2002,Kirman1991}.

\vspace{3mm}

\section{Trends}

As previously said, emergence and recent progress of two separate fields of research will likely have a major upcoming impact on economic and financial ABMs. The first one is the recent developments in cognitive neuroscience and neuroeconomics~\cite{Neuroeconomics2009,Camerer2013,Demartino2013,Camerer2013b,Camerer2013c}, which has revolutionized behavioral economics with its ever lower cost experimental methods of functional magnetic resonance imaging (fMRI), electro-encephalography (EEG), or magneto-encephalography (MEG) applied to decision~\cite{Kahneman1979,Frydman2012} and game theories~\cite{Camerer2003a}. The second one concerns the recent progress of machine learning which has been possible in the last few years by the improved computer power, and especially GPU computing~\cite{Coates2013}, as well as the availability of big data to train supervised algorithms and reach in some tasks superhuman performance~\cite{Wiering2010,Heinrich2015,Silver2016,Mnih2015}. Among the multiple machine learning approaches and algorithms, we can highlight in particular the recent progress of reinforcement learning~\cite{Watkins1992,Sutton1998,Doll2015}, artificial neural networks~\cite{Goodfellow2016,Schmidhuber2015,Turchenko2011}, Monte Carlo tree search~\cite{Silver2016}, and multi-agent learning~\cite{Tuyls2012,Heinrich2014,Heinrich2015}. Furthermore, these have been combined to work together in new types of unsupervised algorithms that have given impressive results~\cite{Reed2016,Lerer2016}, raising concern about the scale of general societal impact of artificial intelligence~\cite{OpenLetter}. 

\vspace{1mm}

To the ABM field, this implies that the realism of the economic agents can be greatly increased via these two recent technological developments: the agents can be endowed with numerous cognitive and behavioral biases emulating those of human investors for instance, but also their trading strategies can be more faithful to reality in the sense that they can be dynamic and versatile depending on the general stochasticity or variability of the market, thanks to the fact that via machine learning they will learn to trade and invest. At a time where the economy and financial markets are progressively more and more automated, this impact of machine learning on ABM should thus be explored. One should keep in mind that the central challenge (and in fact hypotheses) of ABM are the realism of the agents, but also the realism of the economic transactions and interactions between the agents.

\vspace{3mm}

\section{Application}

Besides studying the agents' collective trading interactions with one another, an ABM stock market simulation could be used to probe the following specific fields of study:

\vspace{1mm}

\noindent \textit{Market macrostructure}: one could study how human cognitive and behavioral biases change agents' behaviour at a level of market macrostructure. The two main topics of market macrostructure that would be of interest are naturally all those pertaining to systemic risk~\cite{Systemic2011,Systemic2013} (bubble and crash formation, problem of illiquidity), and also those related to herding-type phenomena~\cite{Bikhchandani2001} (information cascades, rational imitation, reflexivity). A key aspect of this work would be also to carefully calibrate and compare the stylized facts to real financial data (we need to see standardized effects such as volatility clustering, leptokurtic log-returns, fat tails, long memory in absolute returns). Indeed some of these macro-effects have been shown to arise from other agent-based market simulators~\cite{Bikhchandani1992,Samanidou2007}, however it is not yet fully understood how these are impacted by the agents learning process nor their cognitive and behavioral biases such as risk aversion, greed, cooperativity, intertemporality, and the like. 

\vspace{1mm}

\noindent \textit{Price formation}: one could study how these biases change the arbitrage possibilities in the market via their blatant violation of the axioms of the Efficient Market Hypothesis~\cite{Fama1965}. Indeed agent-based market simulators so far have often relied on the use of the aforementioned 'noise traders' in order to generate the necessary conditions for basic business activity~\cite{Sornette2014,Xu2014,Lebaron2002}. The novel aspect here would be to go one step further and replace this notion of purely random trading by implementing specific neuroeconomic biases common to real human behaviors. Another problem of past research in agent-based stock market simulators is that these have often relied on agents getting their information for price forecasting from a board of technical indicators common to all agents~\cite{Pereira2009}. In contrast, we should develop agent models that allow each agent to be autonomous and use and ameliorate its own forecasting tools. This is a crucial aspect in view of the well-known fact that information is at the heart of price formation~\cite{Sanford1980}. We hypothesize that such agent learning dynamics is fundamental in effects based on market reflexivity and impact on price formation, and hints to the role played by fundamental pricing versus technical pricing of assets. Of major interest is the issue of global market liquidity provision and its relation to the law of supply and demand and bid-ask spread formation~\cite{Kyle1985,GrossmanMiller1988,Cason1999}. 
%\todo[inline]{All this directly bears on market microstructure approach!}
\vspace{3mm}

\noindent \textit{Credit risk}: one could study the effect of market evolutionary dynamics when the agents are allowed to learn and improve their trading strategy and cognitive biases via machine learning. In other words, it would be interesting to see the agents population survival rates~\cite{Evstigneev2009,Saichev2010,Malevergne2013} (cf. credit risk), and overall price formation with respect to the arbitrage-free condition of markets when we increase the variability of the intelligence in the trading agents~\cite{Hasanhodzic2011}. Indeed, recent studies~\cite{Sornette2014,Malkiel2012,Black1985} suggest that arbitrage opportunities in markets arise mainly from the collective number of non-optimal trading decisions that shift the prices of assets from their fundamental values via the law of supply and demand. Another topic would be to compare such agent survivability with Zipf's law~\cite{Malevergne2013}. 
%\todo[inline]{This discussion would be more meaningful if we add some illustrations rather than just mention it. And what about other phenomena --- those mentioned below, excess volatility, equity premium puzzle, cascades, bubbles?}
% stating that the number of firms with size greater than $S$ is inversely proportional to $S$. 

%\vspace{3mm}

%\noindent \textit{Price forecasting}: one could gauge the potential of these more realistic ABM simulations to forecast real financial time series via reverse-engineering, a promising new perspective in the field of statistics opened by agent-based models~\cite{Wiesinger2012,Andersen2005,Zhang2013}: the agent-based model parameters are constrained to be calibrated and fit real financial time series and then let to evolve over a given time period as a basic forecast measure on the original time series used for calibration. 

%\vspace{3mm}

%\noindent \textit{Multi-Agent Learning}: It has been shown that a reinforcement learning algorithm could learn to play backgammon~\cite{Wiering2010}, poker~\cite{Heinrich2015}, or go [REF XXX] by just playing against itself and reach super-human capabilities. A promising research perspective of our MAS stock market simulator is to extrapolate the ideas of self-play reinforcement learning (one algorithm learning against itself), and Multi-Agent Learning (many algorithms learning together against themselves) and apply them to trading. 

\vspace{3mm}

\section{Conclusion}

We have thus highlighted new possible exciting perspectives for financial ABM, where the agents would be designed with neuroeconomic biases and having trading or investment strategies updated by machine learning. One should recall that the main argument against ABM, and indeed their main challenge, has always been about the realism of the agents, albeit one should also consider the realism of the economic transactions. We thus argue that these recent trends should set a totally new level of realism for financial ABM
%\todo[inline]{In the light of the above examples, I would limit it to finance}. 

In particular, whereas early financial ABM generations increased their realism of emulation of real stock markets by re-enacting stylized facts gradually during the late nineties, and whereas the issue of calibration is still undergoing the process of automation so that ABM may be validated on large scales of data, we expect these trends to bring in several revolutionary breakthroughs, and the emergence and recognition of ABM as the relevant tools that they are in finance and economics.

Aknowldgement: This work was supported by the RFFI grant nr.16-51-150007 and CNRS PRC nr.151199.

% \clearpage

\medskip

\bibliography{Article_2_}

\end{document}